%% file: main.tex
\documentclass[runningheads]{llncs}
\usepackage[english]{babel}
\usepackage[T1]{fontenc}
\usepackage[utf8]{inputenc}
\usepackage{microtype}

\usepackage{graphicx}          
\usepackage{booktabs}          
\usepackage{tabularx}          
\usepackage{listings}          
\usepackage[table]{xcolor}
\usepackage{amsmath}
\usepackage{enumitem}
\usepackage{tikz}
\usetikzlibrary{positioning,arrows.meta,shapes.geometric,calc,fit,backgrounds}
\usepackage{float}             
\usepackage{hyperref}          
\usepackage{xspace}

\newcommand{\crypistry}{\emph{Crypistry}\xspace}
\newcommand{\crypsy}{\emph{Crypsy}\xspace}
\newcommand{\cryben}{\emph{Cryben}\xspace}

\graphicspath{{./imgs}}

\makeatletter
\newcommand{\includetikzinput}[1]{%
  \begingroup
    \@ifundefined{Ginput@path}{}{%
      \let\input@path\Ginput@path
    }%
    \input{#1}%
  \endgroup
}

\newcommand{\includetikz}[2][\width]{%
  \resizebox{#1}{!}{\input{#2}}%
}

\makeatother

\colorlet{fontblack}{black!85}
\colorlet{bggray}{gray!4}
\colorlet{fggray}{gray!20}
\colorlet{fontgray}{darkgray}
\colorlet{bgblue}{blue!4}
\colorlet{fgblue}{blue!12}
\colorlet{bgorange}{orange!12}
\colorlet{bgred}{red!12}
\colorlet{bgviolet}{red!12!blue!12}
\colorlet{bggreen}{green!6}
\colorlet{bgyellow}{yellow!8}
\colorlet{fgyellow}{yellow!16}

\begin{document}

\title{Hidden Ciphers and Where to Find Them:\\Static Discovery and Assessment of Cryptographic Assets in Software}
\titlerunning{Hidden Ciphers and Where to Find Them}

\author{Christian Näther\inst{1}\orcidID{0009-0000-0634-817X} \and
        Eduard Hirsch\inst{2}\orcidID{0000-0001-9593-2342}}
\authorrunning{C. Näther, E. Hirsch}

\institute{XITASO GmbH, Augsburg, 86153 Germany\\
  \email{christian.naether@xitaso.com}
  \and
  University of Applied Sciences Amberg-Weiden, Amberg, 92224 Germany
  \email{e.hirsch@oth-aw.de}
}




\maketitle

\begin{abstract}
Modern software systems rely on cryptography for data protection, authentication, and trust establishment, yet organizations often lack a structured view of the cryptography deployed across source code, configuration, dependencies, and cryptographic files. This lack of visibility complicates security governance and post-quantum migration planning.

This paper presents a static approach for discovering and assessing cryptographic assets in software systems. We introduce a classification of Crypto-Material, Crypto-Artifacts, and Crypto-Invocations, derive an extensible scanner-independent rule repository from it, and implement a static scanner that applies these rules to produce CBOM-oriented output.

We evaluate the approach on a synthetic benchmark with known ground truth and on a real-world infrastructure of ten deployed services. The scanner achieves an F1 score of 0.75 for asset discovery and correctly annotates 91\% of expected weaknesses and vulnerabilities. In the real-world setting, it processes 57\,610 files in under six minutes and discovers 370 cryptographic assets, including six CVE-linked vulnerabilities and 52 post-quantum migration candidates. Real-world coverage is assessed against a manually compiled reference list rather than an exhaustive one. These results show that classification-driven static discovery can provide practical cryptographic transparency for governance and post-quantum migration planning.
\end{abstract}

\paragraph*{Keywords.}
Cryptographic Transparency $\cdot$ Cryptographic Visibility $\cdot$ Cryptographic Discovery $\cdot$ Static Analysis $\cdot$ CBOM $\cdot$ CWE $\cdot$ CVE $\cdot$ Post-Quantum Cryptography $\cdot$ PQC Migration

\input{sections/01_introduction}
\input{sections/02_related_work}
\input{sections/03_methodology}
\input{sections/04_classification}
\input{sections/05_rule_registry}
\input{sections/06_scanner}
\input{sections/07_evaluation}
\input{sections/08_limitations}
\input{sections/09_conclusion}

\bibliographystyle{splncs04}
\bibliography{bibliography-cleaned}

\appendix

\section{Appendix}
\label{sec:appendix}

\begin{table}[H]
\centering
\caption{Discovery categories for cryptographic assets, with representative subcategories, detected file types, and examples.}
\label{tab:discovery_categories}
\renewcommand{\arraystretch}{1.3}
\setlength{\tabcolsep}{6pt}
\scriptsize
\begin{tabularx}{\columnwidth}{p{13mm} X>{\raggedright\arraybackslash}p{2.4cm} >{\raggedright\arraybackslash}p{2.6cm}}
\toprule
\textbf{Crypto\newline Category} & \textbf{Subcategories} & \textbf{Detected\newline File Types} & \textbf{Examples} \\
\midrule
Material
  & Symmetric and asymmetric keys, SSH and PGP keys, PuTTY and WireGuard keys, Cosign and Minisign keys, JWTs and JWKs, Ansible Vault files, Kerberos credentials, seeds, tokens, nonces, salts, digests, ciphertexts, entropy blobs
  & \texttt{.key}, \texttt{.pem}, \texttt{.ppk}, \texttt{.gpg}, \texttt{.pgp}, \texttt{.jwt}, \texttt{.jwk}, \texttt{.p8}, \texttt{.snk}, \texttt{.keytab}, \texttt{.vault}, \texttt{.secret}, \texttt{.token}, \texttt{id\_rsa}, \texttt{id\_ed25519}, \texttt{cosign.key}
  & RSA private key in \texttt{id\_rsa}; JWT bearer token in \texttt{token.jwt}; GCM nonce in \texttt{nonce.hex} \\
Artifacts
  & X.509 certificates and chains, PKCS\#12 and JKS keystores, CMS and PKCS\#7 containers, certificate signing requests, certificate revocation lists, detached signatures, timestamp tokens, Apple provisioning profiles, KeePass databases, integrity manifests
  & \texttt{.crt}, \texttt{.cer}, \texttt{.der}, \texttt{.pem}, \texttt{.p12}, \texttt{.pfx}, \texttt{.jks}, \texttt{.p7s}, \texttt{.p7b}, \texttt{.csr}, \texttt{.crl}, \texttt{.tsr}, \texttt{.mobileprovision}, \texttt{.kdbx}, \texttt{SHA256SUMS}
  & X.509 leaf certificate in \texttt{.crt}; PKCS\#12 keystore in \texttt{.p12}; SHA-256 integrity manifest \\
Invocations
  & Source-level API calls, TLS and SSH configuration directives, cipher suite selections, certificate references, dependency declarations, provider selections, custom wrappers
  & \texttt{.go}, \texttt{.rb}, \texttt{.py}, \texttt{nginx.conf}, \texttt{sshd\_config}, \texttt{go.mod}, \texttt{Gemfile}, \texttt{requirements.txt}
  & \texttt{"aes.NewCipher(key)"} in Go; \texttt{ssl\_protocols TLSv1.2} in \texttt{nginx.conf}; \texttt{gem "bcrypt"} in \texttt{Gemfile} \\
\bottomrule
\end{tabularx}
\end{table}

\begin{figure}[H]
  \centering
  \includetikz[0.85\textwidth]{extractor_techniques.tex}
  \caption{Extraction techniques used in Phase~2. \emph{Format-Parsing} (left) recovers cryptographic properties from structured file formats. \emph{AST Analysis} (center) matches source-code call patterns against rule-defined signatures. \emph{Regex Matching} (right) extracts cryptographic settings and parameters from configuration files.}
  \label{fig:extractor-techniques}
\end{figure}

\begin{figure}[H]
  \centering
  \includetikz[0.8\textwidth]{crypistry_population_process.tex}
  \caption{Population process for \crypistry, divided into two parallel tracks. \emph{Populate Discovery Rules} identifies target languages and file formats, collects and normalizes library APIs, and derives discovery rules per asset category. \emph{Populate Assessment Rules} reviews CWE and CVE sources, filters by scanner-detectable findings, and derives weakness and vulnerability rules. Both tracks contribute their rules to \crypistry.}
  \label{fig:crypistry-population-process}
\end{figure}

\begin{figure}[H]
  \centering
  \includetikz[0.92\textwidth]{ci_cd_pipelines_details_v2.tex}
  \caption{Research infrastructure used for the real-world evaluation scenario, comprising a GitLab Server, a Runner Server, a Docker Registry, and Deployment Targets. Each service is statically scanned by \crypsy individually, producing one CBOM per service.}
  \label{fig:research_infrastructure}
\end{figure}

\end{document}

%% file: sections/01_introduction.tex
\section{Introduction}
\label{sec:introduction}

Modern software systems use cryptography to protect data, authenticate users, and establish trust across communication protocols, cloud services, software supply chains, and identity management~\cite{nistCryptographicStandards}.
As quantum computing advances and regulatory expectations for cryptographic governance increase, organizations must understand which cryptographic mechanisms are deployed in their software~\cite{nistPQCMigration,nsaCNSA20}. Recent work on post-quantum migration shows that migration efforts remain difficult to structure in practice and depend on reliable knowledge about existing cryptographic assets~\cite{naether2024migratingSoftwareSystemsPQC}. Yet this visibility is often missing, which complicates security assessment, governance, and post-quantum migration planning~\cite{cisa2023quantumreadiness}.

A key reason is that cryptography is scattered across many parts of a software system.
It appears in source files, libraries, configurations, dependency metadata, cryptographic file formats, and binary artifacts~\cite{mousavi2025securityapis,fan2024CryptographicFunctionIdentification}.
Its forms also differ: private keys and seeds are opaque byte sequences, certificates and PKCS\#12 keystores are structured artifacts, and API calls or protocol settings appear as invocations in code and configuration.
Without a common classification, it remains unclear which assets should be discovered, how findings should be grouped, and whether a scanner achieves meaningful coverage~\cite{fan2024CryptographicFunctionIdentification}.

Existing approaches address this problem only partially.
Many tools focus on known misuse patterns, while inventory-oriented tools often couple detection and assessment knowledge tightly to their scanner implementation.
This limits maintainability, extensibility, and adaptation to new asset types, libraries, and security knowledge~\cite{mousavi2025securityapis}.
As a result, cryptographic assets are often found through ad hoc checks rather than through a structured and repeatable discovery process.

To address these challenges, we propose a classification-driven approach for static cryptographic discovery and assessment, combining an asset taxonomy, an extensible rule repository, a static scanner, and an empirical evaluation. The source code, rule repository, benchmark corpus, and evaluation results are available in our supplementary material~\cite{supplementaryMaterials}. Our contributions are:

\begin{itemize}
  \item A classification of cryptographic assets in software, covering cryptographic material, artifacts, and invocations with their subcategories.
  \item An extensible rule repository that organizes detection and assessment rules independently of scanner logic and maps findings to weaknesses and vulnerabilities.
  \item A static rule-based scanner that discovers and assesses cryptographic assets at rest and exports CBOM-oriented results.
  \item An empirical evaluation using a synthetic benchmark with known ground truth and a real-world software system.
\end{itemize}

%% file: sections/02_related_work.tex
\section{Related Work}
\label{sec:related_work}

We discuss related work in four areas: cryptographic API misuse detection, cryptographic inventory and CBOM generation, binary and runtime discovery, and assessment through weakness and vulnerability taxonomies.

\noindent\emph{Cryptographic API Misuse Detection.}
Static analysis is widely used to detect cryptographic API misuses.
Arzt~et~al.~\cite{Arzt2015Towards} showed that secure cryptographic integration requires explicit usage constraints, not only lists of forbidden APIs.
CrySL~\cite{Krueger2021CrySL} captures such constraints as rules over call sequences, predicates, and values.
CryptoGuard~\cite{Rahaman2019CryptoGuard} uses backward program slicing to detect misuse classes in large Java projects, while CRYScanner~\cite{Choudhari2022CRYScanner} shows that many misuses are semantic and relational rather than isolated API calls.
Recent work extends misuse detection to Python~\cite{Frantz2024Cryptolation,Cho2025Cryptbara}, where dynamic language features complicate static reasoning.
These approaches mainly search for known insecure usage patterns.
Our work instead first inventories cryptographic assets and applies weakness and vulnerability assessment afterwards.

\noindent\emph{Cryptographic Inventory and CBOM Generation.}
Cryptographic inventories are increasingly important for governance and post-quantum migration.
Wind River's Crypto Detector~\cite{windriver2020cryptodetector} and the Node.js CBOM approach by Leirimaa~\cite{leirimaa2024supportingPQC} address this need with keyword- and pattern-based discovery, which reports invocation-level evidence but neither parses artifacts nor extracts key material.
CBOMkit~\cite{ibm2025cbomkit} supports CycloneDX CBOM generation, compliance checks, and visualization, but splits discovery across separate engines rather than one classification, so the asset types are not modeled uniformly.
Cryptoscope~\cite{moffie2025Cryptoscope} builds cryptographic inventories by combining source-code analysis with cryptographic domain knowledge, but leaves certificates and keystores to future work.
Schmitt~et~al.~\cite{schmitt2024CriteriaToolingCryptoInventories} define criteria and tooling requirements for cryptographic inventories and propose a prototype covering package-manager and hardware-level detection.
An architecture-derived CBOM approach complements tool-derived inventories by capturing security intent, architectural rationale, and migration-relevant context \cite{hirsch2026architecture}.
These works show the need for crypto inventories, but only partly address the classification question: which asset types should be distinguished for discovery, and how should this guide rule design?
CycloneDX~v1.7 defines how cryptographic assets can be represented in a CBOM, but not how they should be discovered.
Our work addresses this gap by distinguishing Crypto-Material, Crypto-Artifacts, and Crypto-Invocations as first-class categories within one model (Section~\ref{sec:rq1_classification}), yielding one consistent inventory instead of the fragmented or partial coverage above, and by deriving detection and assessment rules from this classification in a scanner-independent repository.

\noindent\emph{Binary, Firmware, and Runtime Discovery.}
Other approaches target cryptography outside source repositories.
Where's Crypto?~\cite{meijer2021whereIsCrypto} identifies proprietary cryptographic primitives in binaries using data-flow graph isomorphism.
CrypTody~\cite{Wang2024CrypTody} detects cryptographic misuse in IoT firmware through data-flow reasoning.
Rattanavipanon~et~al.~\cite{rattanavipanon2025toolchainPQC} propose a binary-level toolchain for post-quantum migration.
Dynamic analysis provides a complementary view: CRYLOGGER~\cite{Piccolboni2020CryLogger} instruments Android apps, records cryptographic API parameters, and validates them offline against rules.
These approaches can observe compiled or runtime behavior, but require binaries, firmware images, or executable systems.
Our work focuses on software systems at rest, including source code, configuration files, dependency metadata, and cryptographic file formats.
This supports deterministic offline scans and CI/CD integration.

\noindent\emph{Assessment Taxonomies and Benchmarks.}
Assessing discovered assets requires a vocabulary for weaknesses and vulnerabilities.
CWE captures recurring cryptographic weakness types, such as risky algorithms, hard-coded keys, and weak pseudo-random number generation.
OWASP A04:2025 Cryptographic Failures provides a broader application-security view.
CVE records describe concrete vulnerabilities in specific products, libraries, or versions.
Many scanners use such taxonomies as their detection scope, yet benchmarks show that advanced misuse detectors still miss relevant finding classes~\cite{Zhang2023AreWeThereYet}.
CamBench~\cite{Schlichtig2022CamBench} further stresses the need for fair and extensible benchmarks for cryptographic API misuse detection.
Our benchmark \cryben has a different goal: it evaluates occurrence-level discovery and CBOM generation across all three asset categories.
Thus, discovery is broader than known weakness classes, while CWE and CVE mappings are applied later to prioritize security-relevant findings.

%% file: sections/03_methodology.tex
\section{Methodology}
\label{sec:methodology}

The development of our approach follows the Design Science Research (DSR) methodology of Hevner~et~al.~\cite{hevner2004}, which focuses on designing and evaluating artifacts for concrete real-world problems. Here, the problem is the lack of a structured approach for identifying, classifying, assessing, and documenting cryptography in software systems. This is relevant for post-quantum migration planning, cryptographic governance, and the detection of insecure primitives.

The outcome is a set of four connected artifacts: a classification of cryptographic assets, the rule repository \crypistry, the static scanner \crypsy, and the synthetic benchmark \cryben. Together, they provide a conceptual and operational foundation for discovering and assessing cryptographic assets at rest.

\begin{figure}[t]
  \centering
  \includetikz[0.95\textwidth]{methodology_dsr_crypsy.tex}
  \caption{Static cryptographic transparency in a Design Science Research setting. The relevance cycle connects practical challenges to the design of the artifacts. The design cycle drives their iterative refinement. The rigor cycle grounds them in scientific foundations. Adapted from Hevner's three-cycle view~\cite{hevner2007threecycle}.}
  \label{fig:methodology}
\end{figure}

Figure~\ref{fig:methodology} positions our work within the \emph{DSR} activity, where the \emph{environment} and the \emph{knowledge base} are linked through the relevance, rigor, and design cycles ~\cite{hevner2007threecycle}.
Within this activity, we develop the four artifacts comprising the classification in Section~\ref{sec:rq1_classification}, the rule repository in Section~\ref{sec:rq2_rule_base}, the scanner in Section~\ref{sec:rq3_scanner}, and the benchmark in Section~\ref{sec:evaluation}.
Each artifact is refined through iterative build-evaluate cycles in the \emph{design cycle}, as reflected in the conceptual design, implementation, and evaluation presented across Sections~\ref{sec:rq1_classification} to ~\ref{sec:discussion}. Two refinement steps illustrate this concretely. First, the classification itself was refined during rule design, which led us to add \emph{Cryptographic Material} as a third category, as detailed in Section~\ref{sec:rq1_classification}. Second, the scanner and \crypistry co-evolved. We began with a small set of discovery rules and a scanner able to detect them, then repeatedly added rules and extended the scanner logic, up to the assessment rules. The classification, the rule repository, and the scanner therefore emerged from repeated build-evaluate steps rather than a single-pass design.

To ground the artifacts in real-world needs, we characterize the application domain using the \emph{environment} activity of Hevner~et~al.~\cite{hevner2007threecycle}, covering the relevant people, organizational context, technical systems, and problems.
In our setting, the environment comprises developers, security engineers, and platform teams who face an organizational need for cryptographic visibility across software systems with distributed cryptographic functionality.

From the \emph{environment} activity, we derive concrete artifact requirements and connect them to design objectives through the \emph{relevance cycle}.
We ground these objectives in the problem of incomplete cryptographic visibility described in Section~\ref{sec:introduction}, operationalize the resulting requirements through RQ1 to RQ4, and validate the artifacts by evaluating \crypsy against a synthetic benchmark \cryben and a real-world software system in Section~\ref{sec:evaluation}.

The \emph{knowledge base} captures the scientific foundations that inform our design, including static analysis, cryptographic inventories and CBOMs, and weakness taxonomies such as CWE and CVE. In the \emph{rigor cycle}, these foundations guide the classification design, detection and normalization, and assessment layers across Sections~\ref{sec:rq1_classification} to~\ref{sec:rq3_scanner}.

Our goal is to provide a structured baseline for crypto inventory generation. We therefore formulate the following research questions to operationalize the design objectives and guide the iterative development and evaluation of the artifacts across the three DSR cycles:

\begin{description}
  \item[RQ1:] \emph{What categories of cryptographic assets can be systematically distinguished in software to support their structured discovery and assessment?}

  With this research question, we define the conceptual foundation of our approach.
  We examine how cryptographic assets in software should be structured for discovery and how discovered assets should subsequently be assessed from a security perspective.
  For discovery, we define the main categories of cryptographic assets that occur in software systems.
  For assessment, we define the categories used to relate discovered assets to weaknesses and vulnerabilities.
  In this way, RQ1 provides the conceptual basis for later rule design and structured assessment.

  \item[RQ2:] \emph{How can the detection and assessment of cryptographic assets be organized in an extensible and maintainable way that keeps domain knowledge separate from scanner logic?}

  RQ2 addresses the organization of detection and assessment knowledge.
  We design \crypistry as a structured repository with discovery-oriented and assessment-oriented rules.
  Assessment rules further distinguish weakness-oriented and vulnerability-oriented concerns.
  We also examine how \crypistry can be populated from cryptographic libraries and external sources such as CWE and CVE DBs.
  The result is a self-contained repository that builds on RQ1 and provides the rule foundation for cryptographic detection and assessment.

  \item[RQ3:] \emph{How can cryptographic assets in software systems be automatically discovered and assessed at rest based on a structured detection and assessment repository?}

  RQ3 operationalizes the repository from RQ2.
  We design and implement \crypsy, a static scanner that loads and applies \crypistry rules for discovery, assessment, and export.
  This realizes the repository in executable form and produces structured, standards-oriented output.

  \item[RQ4:] \emph{To what extent does \crypsy correctly discover and assess cryptographic assets in real-world software systems, and how do the resulting findings distribute across asset categories and target systems?}

  With this research question, we evaluate the practical performance of \crypsy.
  We first examine to what extent the scanner correctly discovers and assesses cryptographic assets against a synthetic benchmark with known ground truth.
  We then study how \crypsy performs in a real-world software system and how the resulting findings distribute across asset categories.
  In this way, RQ4 provides the empirical basis for assessing the utility, quality, and practical limitations of the overall approach.
\end{description}

%% file: sections/04_classification.tex
\section{Classification of Cryptographic Assets}
\label{sec:rq1_classification}

In this section, we define the classification and taxonomy that provide the conceptual foundation for the rest of the paper. Our goal is to establish a clear and operational view of cryptography in software systems so that it can be discovered and assessed in a consistent way. To structure this view, we distinguish between two perspectives: discovery asks what kind of cryptographic element is present and where it occurs, while assessment asks whether a discovered element relates to known weaknesses or vulnerabilities. In the following, we refer to individual elements of cryptography in software as \emph{cryptographic assets}, or \emph{crypto-assets}, to enable more precise categorization and discussion.

\subsection{Cryptographic Asset Categories for Discovery}

We distinguish three categories of statically discoverable cryptographic assets: Value-centric \emph{Cryptographic Material}, \emph{Cryptographic Artifacts} which are structure-centric, and finally context-centric \emph{Cryptographic Invocations}.

\begin{description}
  \item[Cryptographic Material] refers to keying material and secret cryptographic values whose disclosure, modification, or misuse could compromise cryptographic security. It includes raw or encoded inputs to operations such as encryption, signing, authentication, key derivation, or key establishment. The defining property is the security role of the value rather than its file format. Examples include symmetric keys, private keys, relevant public keys, seeds, wrapped keys, authentication tokens, and statically observable session or ephemeral keys.

  \item[Cryptographic Artifacts] describe structured cryptographic objects with security-relevant metadata that can be parsed directly, such as validity periods, issuer or subject names, algorithm identifiers, extensions, trust-store membership, or policy attributes. These objects support trust establishment, policy enforcement, configuration, attestation, validation, or integrity checking. Examples include X.509 certificates, certificate chains, trust anchors, revocation objects, keystores, signature containers, timestamp tokens, signed manifests, and integrity evidence. Containers are classified as artifacts, while extracted secret values are classified as material.

  \item[Cryptographic Invocations] collect observable usage points that select, configure, or call cryptographic functionality in source code, configuration files, build files, dependency manifests, or deployment descriptors. The meaning of these invocations depends on the system context and may reveal algorithms, protocols, modes, parameters, providers, or library versions. Examples include cryptographic API calls, protocol and cipher-suite selections, algorithm identifiers in configurations, library imports, dependency declarations, provider selections, and custom wrapper functions.
\end{description}

We arrived at this classification iteratively while building \crypistry and \crypsy, as described in the design cycle in Section~\ref{sec:methodology}. We initially distinguished only \emph{Artifacts} and \emph{Invocations}, but found during rule design that secret values fit neither, which led us to introduce \emph{Cryptographic Material} as a third category. Each asset is assigned by the role of its content rather than by its format alone. A PEM file holding a certificate is an \emph{Artifact}, whereas a PEM file holding a private key is \emph{Material}. This approach avoids ambiguity and keeps the assignment of discovery rules consistent across categories, so that every discovery rule in \crypistry maps to exactly one of the three categories.

The categories introduced above define the discovery scope of our approach. Table~\ref{tab:discovery_categories} in the Appendix summarizes their representative subcategories and typical evidence forms.

\subsection{Cryptographic Asset Categories for Assessment}

For assessment, the focus shifts from how cryptography appears in software to what security-relevant properties a discovered asset exhibits. The goal is to determine whether a discovered asset relates to known weaknesses or vulnerabilities. Weaknesses are grounded in taxonomies such as CWE. Vulnerabilities are grounded in concrete entries such as CVE. We distinguish two main categories for assessment.

\begin{description}
  \item[Weakness.]
  A weakness describes a security-relevant property of a discovered cryptographic asset.
  Typical examples include broken or risky algorithms, hard-coded keys, insecure random number generation, or unsafe protocol configurations.
  Weaknesses are grounded in taxonomies such as CWE and describe issues at the level of design choices, usage patterns, or configuration properties.

  \item[Vulnerability.]
  A vulnerability describes a concrete and known security issue that affects a discovered cryptographic asset, most often through a specific product, library, component, or version.
  Those include vulnerable dependency versions or known implementation flaws documented in vulnerability databases such as CVE.
  Vulnerabilities therefore capture a more specific and externally grounded assessment view than weaknesses.
\end{description}

\begin{table}[H]
\centering
\caption{Assessment categories for cryptographic assets, grounded in weakness and vulnerability taxonomies.}
\label{tab:assessment_categories}
\renewcommand{\arraystretch}{1.3}
\setlength{\tabcolsep}{8pt}
\begin{tabularx}{\columnwidth}{@{}l X X@{}}
\toprule
\textbf{Category} & \textbf{Typical Subcategories} & \textbf{Examples} \\
\midrule
Weakness
  & Broken or weak algorithm, hard-coded secret, insecure random number generation, unsafe protocol configuration, missing integrity check
  & Use of MD5 for integrity (CWE-327); hard-coded cryptographic key (CWE-321) \\
Vulnerability
  & Vulnerable library version, known implementation flaw in a specific component or version
  & OpenSSL Heartbleed (CVE-2014-0160); OpenSSL padding oracle (CVE-2016-2107) \\
\bottomrule
\end{tabularx}
\end{table}

Table~\ref{tab:assessment_categories} summarizes both categories with typical subcategories and examples. These two assessment categories are not mutually exclusive. A discovered asset may be associated with multiple assessment results. For example, a cryptographic invocation may relate to a CWE-based weakness, while the affected library version may at the same time be linked to a CVE-based vulnerability. Together, the three discovery categories and the two assessment categories answer RQ1: cryptographic assets in software can be systematically distinguished along two orthogonal dimensions, one structural for discovery and one evaluative for assessment.

%% file: sections/05_rule_registry.tex
\section{\crypistry: A Rule-Based Detection and Assessment Repository}
\label{sec:rq2_rule_base}

In this section, we address RQ2 by designing \crypistry, a structured and extensible repository of detection and assessment rules. \crypistry is designed to be maintainable and independent of any scanner implementation, allowing detection and assessment knowledge to evolve without changes to the scanner pipeline. We first describe the structure and rule types, then explain how we populate it.

\subsection{Design and Structure of \crypistry}

\crypistry organizes its rules into two top-level types, as summarized in Figure~\ref{fig:crypistry-rule-structure}. Discovery rules reflect the structural distinction between file-based and invocation-based assets, while assessment rules mirror the weakness and vulnerability categories from Section~\ref{sec:rq1_classification}. We first discuss discovery rules, followed by assessment rules.

\emph{Discovery Rules.} Discovery rules define how cryptographic assets are detected. \crypistry preserves all three discovery categories from Table~\ref{tab:discovery_categories} as distinct subtypes, namely cryptographic material, artifacts, and invocations. Each rule carries a \emph{rule identifier}, a \emph{rule subtype}, an \emph{asset properties} block that pre-classifies the expected \emph{CryptoComponent} type, and a \emph{detector} block that specifies the scan method and file globs used to identify target files. Material rules target files whose type alone unambiguously identifies the content and use filename- and path-based matching, such as a PuTTY private key or a HashiCorp Vault token file. Artifact rules target structured file formats whose contents are ambiguous until parsed and therefore rely on parser-based extraction, such as PEM-encoded files that may contain either a certificate or a private key. Invocation rules cover API calls, configuration settings, and dependency declarations, using AST-based call matching, regex-based matching, or manifest parsing depending on the target. A single rule is not bound to a single file, so a small number of rules achieves broad coverage across many concrete files, formats, and deployment contexts.

\emph{Assessment Rules.} Assessment rules define how discovered and normalized assets are evaluated from a security perspective. \crypistry distinguishes two assessment subtypes, namely weakness-oriented and vulnerability-oriented rules, mirroring the two assessment categories from Table~\ref{tab:assessment_categories}. Each rule carries a \emph{rule identifier}, a severity level, a \emph{mappings} block recording relevant CWE or CVE identifiers, and an \emph{assessor} block specifying the match conditions against a normalized \emph{CryptoComponent}. Because assessment rules operate on already-discovered assets rather than raw files, they require no \emph{detector} block and remain fully independent from extraction logic. For library-based findings, the distinction between weakness and vulnerability rules is particularly relevant, as it differentiates between the mere presence of a cryptographic library and a version known to be vulnerable.

\begin{figure}[t]
  \centering
  \includetikz{crypistry_rule_structure.tex}
  \caption{Representative rules from \crypistry. Discovery rules carry a \emph{detector} block for asset extraction. Assessment rules carry \emph{mappings} and \emph{assessor} blocks relating findings to CWE or CVE identifiers and operate on already-discovered assets.}
  \label{fig:crypistry-rule-structure}
\end{figure}

\subsection{Populating and Extending \crypistry}

\crypistry is populated in two parallel tracks, illustrated in Figure~\ref{fig:crypistry-population-process} in the Appendix, one for discovery rules and one for assessment rules. Both tracks are grounded in the classification from Section~\ref{sec:rq1_classification} and are described in the following.

\emph{Populating Discovery Rules.} For file-based discovery, we aimed to cover all relevant cryptographic file formats across the Material and Artifact categories. This required a deliberate classification. Files with a \emph{.pem} extension, for example, are placed in the Artifact category since they may contain either a certificate or a private key, and the scanner resolves the distinction at parse time rather than by file extension alone. For source-level discovery, we selected Ruby and Go as the first supported languages, as both are widely used in modern infrastructure and no comparable tools exist for these ecosystems. For each language, we identified the most relevant cryptographic libraries, filtered the API calls exposed by their current versions, and derived rules with matching file globs and scan methods as described in Section~\ref{sec:rq2_rule_base}. The same approach applies to configuration-based discovery, surveying relevant server and service formats and deriving rules for cryptographic parameter settings. This results in rules covering the OpenSSL, bcrypt, and jwt-ruby libraries for Ruby, the \emph{Go standard library} and \emph{golang.org/x/crypto} for Go, TLS and SSH configuration formats, Go module and Ruby gem dependency declarations, and file-based discovery across all Material and Artifact categories. In total, \crypistry currently contains 148 discovery rules.

\emph{Populating Assessment Rules.} For assessment, we reviewed crypto-relevant entries in the MITRE CWE database for weakness-oriented rules and the NIST NVD for vulnerability-oriented rules, selecting in each case those findings that a static scanner can reliably detect. Each weakness rule carries a severity level and one or more CWE identifiers. Vulnerability rules cover well-known protocol and certificate attacks as well as library-specific CVEs matched by package name and version range, covering OpenSSL and selected Ruby and Go libraries.

In total, \crypistry currently contains 214 rules, comprising 148 discovery rules and 66 assessment rules.

%% file: sections/06_scanner.tex
\section{Design and Implementation of \crypsy}
\label{sec:rq3_scanner}

In this section, we address RQ3 by designing and implementing \crypsy, a static scanner that loads the rules provided by \crypistry and applies them in a deterministic, four-phase pipeline covering rule loading, asset discovery, asset assessment, and CBOM export. Figure~\ref{fig:crypsy-pipeline-overview} provides a high-level overview. In the following, we describe each phase in order.

\begin{figure}[t]
  \centering
  \includetikz[0.75\textwidth]{crypsy_pipeline_overview.tex}
  \caption{High-level pipeline of \crypsy, consisting of four sequential phases: rule loading, asset discovery, asset assessment, and CBOM export.}
  \label{fig:crypsy-pipeline-overview}
\end{figure}

\emph{Phase 1: Rule Loading.} \crypsy begins by loading the selected ruleset from \crypistry into two structured collections, namely discovery rules that drive Phase~2 and assessment rules that drive Phase~3.
The \emph{default} ruleset covers all supported asset categories and language ecosystems, while more focused rulesets restrict the scan to a specific asset category or programming language.
Once files have been collected and classified in Phase~2, rules whose target category is absent from the scan are automatically deactivated, keeping the active rule set focused without requiring manual configuration.
Both collections are passed through the remaining pipeline as an immutable rule set, ensuring that every phase operates on the same consistent rule state.

\emph{Phase 2: Discovery.} The discovery phase proceeds in four steps covering file collection, file classification, extraction, and normalization.
\crypsy collects all files within the target scope and classifies each file into a discovery category.
Classification for Crypto-Material and Crypto-Artifacts is driven entirely by the \emph{file globs} declared in the loaded discovery rules, so new material and artifact formats are automatically recognized without any change to the scanner code.
Source code, configuration files, and dependency manifests are classified by their known file extensions and filenames, as these categories are not expressed as file-parsing rules in \crypistry.
Each discovery category relies on a distinct extraction technique, as illustrated in Figure~\ref{fig:extractor-techniques} in the Appendix, where Crypto-Material and Crypto-Artifacts are processed through format-based parsing and Crypto-Invocations are processed through AST analysis for source-level calls, regex matching for configuration files, and manifest parsing for dependency declarations.
Each extractor produces raw findings that are normalized into canonical \emph{CryptoComponent} objects, providing a uniform representation for all discovered asset types across Phase~3.

\emph{Phase 3: Assessment.} The assessment phase evaluates each \emph{CryptoComponent} discovered in Phase~2 against the assessment rules from \crypistry, producing results with severity and mappings to CWE or CVE identifiers.
Each component is checked against both weakness rules and vulnerability rules.
Weakness rules flag security-relevant properties such as weak algorithms or insecure configurations, while vulnerability rules match components against known vulnerabilities in specific versions or protocol implementations.
A TLS~1.0 configuration, for instance, may trigger both a weakness rule for the deprecated protocol and a vulnerability rule for the BEAST attack.
Similarly, a library dependency may trigger a vulnerability rule when its version falls within a known vulnerable range, as in the case of \emph{openssl} versions affected by Heartbleed (CVE-2014-0160).

\emph{Phase 4: Export.} In the export phase, \crypsy maps its internal asset model to a target schema and produces a structured output file.
The export phase is intentionally decoupled from the preceding pipeline phases, so additional schemas can be added without affecting discovery, assessment, or normalization.
The current implementation supports CycloneDX~v1.7 CBOM, capturing discovered assets together with their properties, evidence locations, and assessment results.
This standards-oriented representation is directly suitable for cryptographic inventorying, governance, and post-quantum migration planning.

\emph{Overall Interaction with \crypistry.} Figure~\ref{fig:crypistry-crypsy-overview} illustrates the overall interaction between \crypistry and \crypsy.
The two artifacts are designed to evolve independently, where \crypistry owns all cryptographic knowledge in the form of discovery and assessment rules and \crypsy owns the pipeline logic that loads and applies them.
This separation ensures that detection and assessment behavior can be extended or updated entirely through changes to \crypistry, keeping the overall approach maintainable as new asset types, languages, and security knowledge emerge.
\crypistry can therefore be versioned, shared, and updated independently, allowing teams to maintain their own rule sets or adopt community-contributed rules without replacing the scanner.

\begin{figure}[t]
  \centering
  \includetikz[0.90\textwidth]{crypistry_crypsy_overview.tex}
  \caption{Overall interaction between \crypistry and \crypsy. \crypistry provides the rules for discovery and assessment, whereas \crypsy applies them across its discovery, normalization, assessment, and CBOM export pipeline.}
  \label{fig:crypistry-crypsy-overview}
\end{figure}

%% file: sections/07_evaluation.tex
\section{Evaluation}
\label{sec:evaluation}

In this section, we address RQ4 by evaluating \crypsy in a benchmark scenario and a real-world scenario.
The benchmark scenario evaluates correctness by running \crypsy against \cryben, a synthetic benchmark corpus with known ground truth, reporting precision, recall, and F1 for discovery and assessment.
The real-world scenario assesses applicability by running \crypsy on a self-operated IT infrastructure of ten deployed services, reporting discovery coverage, assessment quality, and post-quantum migration candidates.

\subsection{Benchmark Evaluation}
\label{sec:eval_benchmark}

In this section, we present the benchmark evaluation of \crypsy against \cryben and report correctness results at the occurrence level.

\emph{Benchmark Corpus.} \cryben is a synthetic benchmark corpus designed to evaluate the correctness of cryptographic asset discovery and assessment, covering all three discovery categories and their corresponding assessment annotations.
The corpus covers Crypto-Material, Crypto-Artifacts, and Crypto-Invocations in Go, Ruby, and nginx configuration, including all CBOM cryptographic asset types defined in CycloneDX~v1.7.
Its ground truth is a CycloneDX~v1.7 CBOM constructed independently of \crypsy, so that any undetected asset constitutes a genuine false negative.

\emph{Discovery Results.} We now report the results of running \crypsy against \cryben, starting with discovery correctness across all three categories, as illustrated in Table~\ref{tab:correctness_by_category}.
Overall, \crypsy achieves an F1 score of 0.75 (Precision\,=\,0.87, Recall\,=\,0.66) across all 197 ground-truth occurrences.
Crypto-Material achieves near-perfect recall (F1\,=\,0.94) because filename-pattern detection is unambiguous, with the two false positives being format variants not yet covered by material rules (a raw Base64-encoded AES key and an EC key in SEC1 format).
Go invocations perform strongly (F1\,=\,0.92), while configuration-level invocations (F1\,=\,0.78) show overlapping rule matches where a single \emph{ssl\_protocols} directive triggers separate rules for each referenced TLS version simultaneously.
Crypto-Artifacts reaches F1\,=\,0.61 with zero false positives, where the two false negatives reflect coverage gaps for uncommon PEM label variants and one case where the scanner correctly merged a duplicate asset into a single component that the ground truth counted twice.
Ruby invocations show the lowest recall (F1\,=\,0.51) because \crypistry targets only the constructor call \emph{OpenSSL::Cipher.new} as the canonical anchor, while the ground truth tracks individual method calls on cipher objects as separate occurrences.
The 20 false positives all correspond to locations with genuine cryptographic relevance, covering overlapping rule matches for the same directive, assets not captured in the ground truth, and co-located findings the evaluator does not match.
Of the 67 false negatives, 43 are by-design counting differences and 24 are true detection gaps. The by-design cases are Ruby occurrences that the scanner already covers once per cipher object, together with overlapping config-rule matches, while the true gaps are files the scanner misses entirely, plus a few Go, artifact, and material cases. Table~\ref{tab:correctness_by_category} reports both views in the summary rows, and the full per-category decomposition is given in the supplementary material. Excluding the by-design differences, recall rises to 0.84 and F1 to 0.86, and at component granularity only 2 of 48 components are missed (recall 0.96, F1 0.95). This shows that most of the headline gap is a counting-model mismatch rather than missed detections.

\emph{Assessment Results.} We now report assessment correctness, examining how well \crypsy annotates discovered components with CWE and CVE identifiers.
Of the 104 discovered components, 21 out of 23 expected identifiers are correctly annotated, yielding an assessment recall of 0.91 and an assessment F1 of 0.74.
The 13 assessment false positives reflect rules that fire on components for which the ground truth does not define an expected identifier.
The two missed annotations concern the SHA-1 certificate component, for which \crypistry assessment rules do not yet cover the \emph{certificate} asset type.
Figure~\ref{fig:assessment_coverage} shows assessment results at component granularity, with correct annotations spanning symmetric cipher weaknesses, hash weaknesses, protocol vulnerabilities, and the library vulnerability CVE-2019-11840 via \emph{go.mod}.

\emph{Comparison with an Existing Tool.} To position \crypsy against an existing tool, we ran CBOMkit-hyperion (the PQCA \emph{sonar-cryptography} plugin, v1.6.1) on the Go-invocation subset of \cryben, which the baseline is designed to detect, and scored both tools with the same occurrence-level evaluator and ground truth. On the full Go-invocation ground truth \crypsy reaches F1\,=\,0.92 (P\,=\,0.95, R\,=\,0.89) versus 0.66 for CBOMkit (P\,=\,0.84, R\,=\,0.54), and it still leads within CBOMkit's own documented coverage (the Go standard library excluding \emph{crypto/x509}, F1\,=\,0.83 vs. 0.72), as summarized in Table~\ref{tab:baseline_comparison}.
The gap is primarily driven by recall. CBOMkit produces no detections at non-cryptographic locations; its apparent false positives result exclusively from reporting more component labels at a correctly identified line than are recorded in the ground truth.The full setup, per-scope results, and a per-call breakdown are provided in the supplementary material.

\begin{table}[ht]
\centering
\caption{Head-to-head occurrence-level comparison of \crypsy and CBOMkit-hyperion (v1.6.1) on the Go-invocation subset of \cryben, scored with the same evaluator and ground truth. \emph{All Go} covers the full Go invocation ground truth, whereas \emph{Go stdlib} restricts to CBOMkit's own documented coverage (Go standard library excluding \emph{crypto/x509}).}
\label{tab:baseline_comparison}
\renewcommand{\arraystretch}{1.3}
\setlength{\tabcolsep}{5pt}
\scriptsize
\begin{tabularx}{\textwidth}{X l c c c c c c c}
\toprule
\textbf{Scope} & \textbf{Tool} & \textbf{Truth} & \textbf{TP} & \textbf{FP} & \textbf{FN} & \textbf{P} & \textbf{R} & \textbf{F1} \\
\midrule
All Go    & \crypsy  & 70 & 62 &  3 &  8 & 0.95 & 0.89 & \textbf{0.92} \\
All Go    & CBOMkit  & 70 & 38 &  7 & 32 & 0.84 & 0.54 & \textbf{0.66} \\
\midrule
Go stdlib & \crypsy  & 55 & 50 & 15 &  5 & 0.77 & 0.91 & \textbf{0.83} \\
Go stdlib & CBOMkit  & 55 & 36 &  9 & 19 & 0.80 & 0.65 & \textbf{0.72} \\
\bottomrule
\end{tabularx}
\end{table}

\begin{table}[ht]
\centering
\caption{Occurrence-level \crypsy correctness on \cryben by discovery category across 197 ground-truth assets, reporting precision (P), recall (R), and F1. The two summary rows report performance excluding the 43 by-design counting differences and at component granularity; the full per-category decomposition is given in the supplementary material.}
\label{tab:correctness_by_category}
\renewcommand{\arraystretch}{1.3}
\setlength{\tabcolsep}{5pt}
\scriptsize
\begin{tabularx}{\textwidth}{X c c c c c c c c}
\toprule
\textbf{Crypto Category} & \textbf{Truth} & \textbf{Scan} & \textbf{TP} & \textbf{FP} & \textbf{FN} & \textbf{P} & \textbf{R} & \textbf{F1} \\
\midrule
Material              & 17 & 19 & 17 &  2 &  0 & 0.89 & 1.00 & 0.94 \\
Artifacts             & 14 &  9 &  7 &  0 &  2 & 0.78 & 0.50 & 0.61 \\
Invocations (Ruby)    & 78 & 39 & 30 &  9 & 48 & 0.77 & 0.38 & 0.51 \\
Invocations (Go)      & 70 & 65 & 62 &  3 &  8 & 0.95 & 0.89 & 0.92 \\
Invocations (Config.) & 18 & 18 & 14 &  6 &  9 & 0.78 & 0.78 & 0.78 \\
\midrule
$\boldsymbol{\sum}$ \textbf{Findings} & \textbf{197} & \textbf{150} & \textbf{130} & \textbf{20} & \textbf{67} & \textbf{0.87} & \textbf{0.66} & \textbf{0.75} \\
$\sum$ \emph{excl.\ by-design} & 197 & 150 & 130 & 20 & 24 & 0.87 & 0.84 & 0.86 \\
$\sum$ \emph{component level}  &  48 &  49 &  46 &  3 &  2 & 0.94 & 0.96 & 0.95 \\
\bottomrule
\end{tabularx}
\end{table}

\begin{figure}[ht]
  \centering
  \includetikz[0.75\textwidth]{assessment_coverage_matrix.tex}
  \caption{Component-level assessment coverage of \crypsy on \emph{Cryben}.
  Rows correspond to ground-truth-relevant components, grouped by discovery category.
  Columns correspond to expected CWE and CVE identifiers, separated into weakness (left) and vulnerability (right).
  Filled circles~($\bullet$) indicate correctly produced annotations~(TP);
  open circles~($\circ$) indicate expected identifiers missed by the current assessment rules~(FN).
  Blank cells indicate that the identifier is not applicable to the component.}
  \label{fig:assessment_coverage}
\end{figure}

\subsection{Real-World Evaluation}
\label{sec:eval_realworld}

In this section, we report the results of running \crypsy on a self-operated IT infrastructure of ten deployed services, covering discovery coverage, assessment quality, and post-quantum migration candidates.

\emph{Research Infrastructure.} The infrastructure comprises ten services spanning three systems, including GitLab services, a CI/CD runner, and additional services such as Mastodon, Harbor, nginx, Redis, and PostgreSQL, as illustrated in Figure~\ref{fig:research_infrastructure} in the Appendix.
The services combine Ruby on Rails and Go codebases with production TLS configurations, deployment secrets, and dependency manifests.
We scan each service as a single unit, processing source code, configuration files, and dependency manifests exactly as they are present in the deployed repository.
Prior to scanning, we conducted a manual inspection of the infrastructure to identify and record existing cryptographic assets as a partial ground truth, covering all three discovery categories.

\emph{Discovery and Assessment Results.} We now report the results of running \crypsy against the real-world infrastructure, starting with the distribution of findings across discovery categories and services as shown in Table~\ref{tab:rw_distribution}.
Cells marked with \emph{--} indicate that the category is not applicable to the service, reflecting the language and file type composition of each service.
GitLab~CE dominates in Ruby invocations due to its large Rails codebase, while nginx contributes exclusively configuration-level findings.
Redis and PostgreSQL contribute only certificate, key material, and configuration-level findings, confirming that the scanner produces meaningful results even for services where source-level invocations are absent.

\begin{table}[ht]
\centering
\caption{Distribution of \crypsy findings by discovery category and service in the real-world infrastructure scan. Cells report normalized CryptoComponents; ``--'' denotes non-applicable categories. $\sum$ aggregates each category across services. Below the rule, weaknesses and vulnerabilities report annotated component counts; scanned files and duration give per-service scan statistics.}
\label{tab:rw_distribution}
\renewcommand{\arraystretch}{1.3}
\setlength{\tabcolsep}{4pt}
\scriptsize
\begin{tabularx}{\textwidth}{p{1.8cm}c>{\centering}p{5mm}>{\centering}>{\centering}>{\centering}p{5mm}>{\centering}p{5mm}>{\centering}p{6mm}
                            >{\centering\columncolor{fgyellow}}p{8mm}>{\centering\columncolor{fgyellow}}p{7mm}>{\centering\columncolor{fgyellow}}c>{\columncolor{fgyellow}}c>{\centering\columncolor{fgyellow}}c>{\bfseries}r}
\toprule
\textbf{}
  & \multicolumn{5}{c}{\textbf{Gitlab}}
  & \multicolumn{5}{c}{\cellcolor{fgyellow}\textbf{External Services}}
  &\\
\textbf{Crypto\newline Category}
  & \textbf{CE}
  & \textbf{Git- aly}
  & \textbf{Ag- ent}
  & \textbf{Pa- ges}
  & \textbf{Run- er}
  & \textbf{Masto- don}
  & \textbf{Har- bour}
  & \textbf{nginx}
  & \textbf{Redis}
  & \textbf{PG}
  & $\boldsymbol{\sum}$ \\
\midrule
Material              &  5 &  6 & -- &  1 &  1 &  2 &  3 &  4 &  1 & 1 &  24 \\
Artifacts             &  9 &  7 & -- &  7 &  3 &  3 & 10 &  7 & -- & 2 &  48 \\
Inv.\ (Ruby)          & 89 & -- & -- & -- & -- & 26 & -- & -- & -- & -- & 115 \\
Inv.\ (Go)            & 17 & 21 & 20 &  9 & 27 & -- & 27 & -- & -- & -- & 121 \\
Inv.\ (Config.)       & 11 &  7 &  1 &  1 &  1 &  9 &  8 & 19 &  4 & 1 &  62 \\
\midrule
$\boldsymbol{\sum}$ \textbf{Findings}
  & \textbf{131} & \textbf{41} & \textbf{21} & \textbf{18} & \textbf{32} & \textbf{40} & \textbf{48} & \textbf{30} & \textbf{5} & \textbf{4} & \textbf{370} \\
\bottomrule
\textit{Weaknesses}  &  31 & 16 &  5 &  1 & 11 &  9 & 12 &  3 & 2 & 1 &  91 \\
\textit{Vulnerabilities} &   2 &  2 &  0 &  0 &  1 &  0 &  0 &  1 & 0 & 0 &   6 \\
\midrule
\textit{Scanned Files}   & 40.0k & 1.5k & 1.3k & 349 & 1.3k & 3.2k & 3.3k & 574 & 843 & 5.4k & 57.6k \\
\textit{Duration (s)} &  207 &   12 &    8 &   2 &   12 &   88 &   21 &   4 &   1 &    2 &   357 \\
\bottomrule
\end{tabularx}
\end{table}

All assets identified during manual inspection were detected, yielding a partial recall of 100\%. This recall of 100\% reflects agreement with our manually built partial ground truth, not true recall against an exhaustive inventory. It should be read as ``no manually identified asset was missed'', not as complete coverage.
Across ten services and 57\,610 files, \crypsy discovered 370 cryptographic components in under six minutes.
A manual review of all 370 components confirmed no incorrect detections, with 43 components originating from test fixture paths that can be excluded through targeted \emph{--exclude} patterns without affecting any production-relevant finding.
Of the 370 components, 91 received at least one CWE-mapped weakness annotation and 6 received at least one CVE-linked vulnerability annotation.
The most frequently triggered weaknesses were insecure PRNG usage (CWE-338), disabled TLS certificate validation (CWE-295), and weak cipher suite configuration (CWE-326). Because these rules do not track how a primitive is used, not every annotation is security-actionable, so we manually reviewed the weakness-annotated components using their recorded location and snippet. Posture weaknesses, such as disabled certificate validation, obsolete TLS versions, or weak cipher suites, are actionable whenever they occur, whereas weak-hash and weak-PRNG annotations depend on the use site and mostly fire on non-security digests, for example a SHA-256 or SHA-1 hash of a file path, cache key, or identifier rather than a password. The actionable weaknesses are therefore dominated by the posture rules, at an actionable-weakness precision of roughly 0.3.
On the vulnerability side, dependency rules identified a vulnerable bcrypt version in GitLab~CE, a Go cryptography library flaw in Gitaly, SSH Logjam in Gitaly and GitLab~Runner, and the BEAST vulnerability (CVE-2011-3389) in the legacy TLS~1.0 nginx configuration, confirming end-to-end coverage from dependency manifests to deployment configuration.

\emph{Post-Quantum Migration.} A review of the CBOM output identifies 52 components as post-quantum migration candidates, comprising RSA and elliptic-curve keys, certificates, and TLS configurations spread across all ten services.
All identified assets rely on algorithms vulnerable under Shor's algorithm~\cite{shor1994algorithms} and would require replacement with NIST post-quantum standards such as ML-KEM, ML-DSA, or SLH-DSA.
The CBOM provides the full evidence chain for each finding, giving security teams the precise file locations and algorithm parameters.
The resulting CBOM serves as a starting point for a structured post-quantum migration, representing a part of the cryptographic inventory that security teams can use to plan and prioritize the necessary algorithm replacements.

Concluding, we note, that Crypsy does not compute migration strategies, but its CBOM output provides asset-level evidence for later migration planning, which recent work frames as a structurally complex dependency problem~\cite{loebenberger2024formalization}.

%% file: sections/08_limitations.tex
\section{Limitations and Threats to Validity}
\label{sec:discussion}

In this section, we discuss the limitations of our approach and reflect on threats to the validity of our evaluation.

\emph{Limitations of \crypistry.} The coverage and correctness of the scanner depend directly on the quality and completeness of \crypistry, which requires ongoing maintenance as cryptographic libraries and usage patterns evolve.
Without dataflow analysis, assessment rules such as \emph{weak.hash.password.weak} fire on all matching invocations regardless of usage context, as the scanner cannot trace how the result is used, requiring manual triage to distinguish actionable from non-actionable findings.
PQC functionality is not yet uniformly available across relevant Ruby cryptographic libraries, so concrete PQC-related function calls cannot yet be covered systematically.
\crypistry currently targets the active API surface of supported libraries, with coverage of historical and deprecated function versions planned as a next extension step.
Support is currently focused on Ruby and Go as the primary language ecosystems, with proprietary and vendor-specific configuration formats not yet covered.

\emph{Limitations of \crypsy.} \crypsy reliably captures calls where algorithm identifiers appear as literals at the call site, but cannot resolve arguments computed at runtime.
On the performance side, the scanner currently processes files sequentially, which was sufficient given the scan times observed in our evaluation but may become a bottleneck for very large codebases.
A related usability limitation is that \crypsy currently requires the user to configure the scan scope manually, as it does not yet provide an automated discovery mode that enumerates installed components and languages before scanning.

\emph{Threats to Validity.}
The evaluation focuses on Ruby and Go, improving internal consistency but limiting generalizability to other languages and contexts. While the classification, the rule repository, and the pipeline are language-agnostic by design, our empirical evidence covers only Ruby, Go, and selected configuration formats. Ecosystems that hold much enterprise cryptography, such as Java, C/C++, and Python, are not yet demonstrated. We therefore scope the ``practical cryptographic transparency'' claim to the evaluated ecosystems and treat broader generalizability as a design expectation to be validated. Extending language coverage is our primary next step.
\cryben was built independently of \crypsy's current rule set, so undetected assets are genuine false negatives caused by scope decisions, \crypistry coverage gaps, or format boundary cases.
For the real-world system, exhaustive ground truth is unavailable; we therefore rely on partial ground truth and manual inspection.

%% file: sections/09_conclusion.tex
\section{Conclusion}
\label{sec:conclusion}

This paper presents a structured approach to statically discover and assess cryptography in software, combining unified classification, extensible rule repository, and static scanner.
We validate the approach against a synthetic benchmark with known ground truth and a real-world infrastructure of ten deployed services.

Within RQ1, we established a classification distinguishing three main discovery categories, namely cryptographic material, artifacts, and invocations, and two assessment categories grounded in weaknesses and vulnerabilities.
Building on this foundation, we designed \crypistry, an extensible rule repository that organizes detection and assessment rules independently of scanner logic (RQ2), and implemented \crypsy, a static scanner that applies these rules to discover and assess cryptography at rest, producing CycloneDX-conformant CBOM output (RQ3).
On a synthetic benchmark with known ground truth, \crypsy achieves an F1 score of 0.75 for asset discovery and correctly annotates 91\% of expected weaknesses and vulnerabilities (RQ4).
Applied to a real-world infrastructure of ten deployed services, it discovers 370 cryptographic assets in under six minutes, including six CVE-linked vulnerabilities and 52 components requiring post-quantum migration.

A further direction is to connect CBOM-based inventories with cryptographic agility processes. Recent work shows that cryptographic agility is still defined inconsistently, suggesting that explicit asset inventories may help operationalize agility in practice~\cite{nather2024toward}.

Taken together, these results demonstrate that rule-based static discovery, grounded in a structured classification, can provide practical cryptographic transparency for deployed software systems. The evidence covers Ruby, Go, and common configuration formats, and the real-world coverage is measured against a partial manual ground truth, so extending language coverage and validating against fuller ground truth remain future work. This gives organizations a starting point to understand their deployed cryptography and plan their post-quantum migration.